\documentclass[aps,prb,reprint,superscriptaddress,longbibliography]{revtex4-2}

\usepackage[T1]{fontenc}
\usepackage[utf8]{inputenc}
\usepackage{amsmath,amssymb,amsfonts,mathtools,bm}
\usepackage{physics}
\usepackage{graphicx}
\usepackage[colorlinks=true,allcolors=blue]{hyperref}
\usepackage[nameinlink,capitalize]{cleveref}
\usepackage{microtype}
\usepackage{listings}
\usepackage{algorithm}
\usepackage{algpseudocode}
\usepackage{xcolor}
\usepackage{bbm}

\newcommand{\Acal}{\mathcal{A}}

\newcommand{\kvec}{\mathbf{k}}

\graphicspath{{./IMG/}}

\begin{document}

\title{A basis-free, octonionic criterion for Weyl points in solids}

\author{Christian Tantardini*}
\affiliation{Center for Integrative Petroleum Research, King Fahd University of Petroleum and Minerals, Dhahran 31261, Saudi Arabia.}

\email{christiantantardini@ymail.com}

\date{\today}

\begin{abstract}
Standard ways of locating and characterizing Weyl points—Berry‐flux integrals on small spheres and local $k\!\cdot\!p$ fits—are topologically sound but depend on user choices (sphere center/radius, gauge smoothing, surface charting, and Pauli‐frame transport) that complicate high‐throughput workflows. We introduce a \emph{local, basis‐free} diagnostic that avoids these knobs by using the octonionic geometry of $\Im\mathbb O\!\cong\!\mathbb R^7$. From a smooth two-band projector we build a unit octonion field $u(\mathbf k)$ and its octonionic connection $\Acal_i=\Im(\bar u\,\partial_{k_i}u)$. Contracting the three directional derivatives with the $\mathrm{G}_2$–invariant three-form produces a pseudoscalar density whose sign gives the node’s chirality. A companion quantity—the octonionic \emph{associator}—vanishes at leading order precisely when the local three directions close inside an associative (quaternionic) three-plane; its smallness thus provides an intrinsic, pointwise self-consistency check that the two-band reduction is valid. The construction is invariant under $\mathrm{SU}(2)$ gauge changes of the two-band subspace and under $\mathrm{G}_2$ rotations of its completion, and it eliminates enclosing surfaces and gauge seams. In the linear regime the density reduces to the oriented volume set by the velocity matrix, thereby agreeing with the Chern number on a small sphere and with the chirality obtained from a $k\!\cdot\!p$ linearization. We outline a practical stencil-based algorithm compatible with Wannier tight-binding Hamiltonians and demonstrate robustness to benign numerical choices, while providing an intrinsic warning signal near band entanglement or multi-fold touchings.
\end{abstract}

\maketitle

\section{Introduction}
\label{sec:I}

Weyl semimetals host point degeneracies in three dimensions at which two nondegenerate bands cross linearly. Each node behaves as a monopole of Berry curvature in momentum space, endowing the band structure with a quantized topological charge (chirality) and giving rise to a range of unusual responses.\cite{Nielsen1981a,Nielsen1981b,Hosur2013CRP,Armitage2018RMP,YanFelser2017ARCMP} On open boundaries, the bulk monopole charge is encoded by surface Fermi arcs that connect projections of opposite chiralities.\cite{Wan2011PRB,Turner2013} In transport, the axial (chiral) anomaly under parallel electric and magnetic fields produces a characteristic negative longitudinal magnetoresistance and related nonlocal effects.\cite{Son2013PRB,Xiong2015Science}
Long before electronic Weyl semimetals were realized, the Adler--Bell--Jackiw chiral anomaly was observed in the chiral superfluid $^3$He-A.\cite{Bevan1997Nature} There the positions of the two Weyl nodes are $\pm p_F\,\hat{\boldsymbol{\ell}}(\mathbf r,t)$, so the texture $p_F\hat{\boldsymbol{\ell}}$ plays the role of an emergent \emph{axial} gauge potential. Motion of a continuous vortex (skyrmion) generates effective axial electric and magnetic fields, drives spectral flow, and produces an anomalous force on the skyrmion with the correct ABJ prefactor---equal to the number of Weyl species (Berry-monopole charges) in $^3$He-A.\cite{Bevan1997Nature}
Optical and nonlinear probes further reveal gyrotropic and circular photogalvanic responses traceable to the Berry curvature texture near the nodes.\cite{Ma2017NatPhys,deJuan2017NatComm}

Following early theoretical proposals,\cite{Wan2011PRB,Burkov2011PRL} \emph{ab initio} predictions and ARPES measurements established the TaAs family as the first realization of a time-reversal--invariant Weyl semimetal,\cite{Weng2015PRX,Xu2015Science,Lv2015PRX,Yang2015NatPhys} with many subsequent materials platforms now known, including systems with strongly tilted (type-II) cones.\cite{Soluyanov2015Nature,Armitage2018RMP} As the materials landscape has expanded, so too has the practical need for robust, automatable diagnostics of Weyl nodes and their chiralities within high-throughput workflows.

In practice, two complementary diagnostics are standard. The first integrates the Berry curvature over a small closed surface (typically a sphere) surrounding a candidate point to obtain the Chern number of the lower band.\cite{Fukui2005JPSJ,Yu2011PRB,Soluyanov2012PRB,Armitage2018RMP,Gresch2017PRB,Wu2018CPC} The second linearizes the two-band Hamiltonian near the crossing, $H_{\rm eff}(\mathbf q)\approx \bm w\!\cdot\!\mathbf q\,\sigma_0+\sum_{ij}v_{ij}q_j\sigma_i$, and reads the chirality from $\mathrm{sgn}\det v$.\cite{Wan2011PRB,Burkov2011PRL,Armitage2018RMP} Both routes are topologically sound in the continuum limit, yet both introduce \emph{algorithmic} choices at finite resolution: selection and smoothing of the two-band subspace in the presence of nearby bands; gauge fixing on an enclosing surface (seams, charts, and phase alignment); the center and radius of the sphere; the discretization and loop orientation on $S^2$; and the choice of local Pauli frame and its transport away from $\kvec_\star$ when taking $k$-derivatives. Additional bookkeeping enters through reciprocal-basis handedness and Brillouin-zone folding in supercell calculations. These choices are benign when carefully controlled, but they complicate high-throughput scans and can seed apparent discrepancies between Berry-sphere and $k\!\cdot\!p$ checks, especially in the presence of tilt, strain, weak disorder, or mild band entanglement.

In this work we formulate a \emph{local, basis-free} diagnostic that reproduces the conventional chirality while bypassing enclosing surfaces and Pauli-frame transport. The construction uses the exceptional algebra of octonions. The imaginary octonions $\Im\mathbb O\cong\mathbb R^7$ carry a canonical $\mathrm{G}_2$--invariant three-form $\varphi$ and a nontrivial associator $[x,y,z]=(xy)z-x(yz)$ that vanishes precisely on associative (quaternionic) three-planes.\cite{Baez2002BullAMS,Bryant1987} We leverage this structure to encode the local two-band geometry into a unit octonion field and define an \emph{octonionic} pseudoscalar density by contracting three directional derivatives with $\varphi$. At a simple Weyl node this density has the sign of $\det v$, while the octonionic associator furnishes an intrinsic self-consistency check: it vanishes to leading order if and only if the local three-direction data close inside a quaternionic subalgebra. In this way, the same object that yields chirality also signals when band entanglement or frame inconsistencies contaminate the two-band reduction.

Our \emph{Octonionic Weyl Point Criterion} (OWPC) thus collapses the earlier knobs to a pointwise computation that is invariant under local $\mathrm{SU}(2)$ gauge changes of the two-band subspace and under $\mathrm{G}_2$ rotations of its completion in $\Im\mathbb O$. The only remaining convention is the global orientation of momentum space, under which any chirality necessarily flips sign. We provide a concrete numerical recipe compatible with Wannier tight-binding Hamiltonians, together with self-consistency checks based on stencil refinement and the associator norm, and we prove equivalence with the conventional Chern-charge and $k\!\cdot\!p$ diagnostics in the linear regime. Beyond surface- and $k\!\cdot\!p$--based diagnostics, there is by now a substantial body of work that casts Weyl nodes and their motion in terms of \emph{quaternion charges} and non-Abelian braiding in parameter space, with applications ranging from generic settings to magnetic and Floquet systems.\cite{RefPRB2020Quat,RefNatPhys2021,RefPRB2021Mag,RefNatComm2024Floquet} In parallel, an exhaustive K-theoretic classification of Weyl nodes (and related band-touching singularities) has been developed, including a clean map to equivariant $K$-theory.\cite{RefPRX2017K} Our contribution is complementary: we do not propose a new classification or braiding framework. Instead, we give a \emph{single-point, basis-free} \emph{diagnostic}---built from octonionic geometry---that (i) reproduces the conventional chirality in the linear regime and (ii) adds an intrinsic self-consistency check via the octonionic associator, aimed specifically at reducing algorithmic arbitrariness in day-to-day computations.

The remainder of the paper is organized as follows. Section~\ref{sec:conventional-human} reviews the conventional workflow and its algorithmic arbitrariness. Section~\ref{sec:owpc} introduces the octonionic connection and $\mathrm{G}_2$ volume density, states the OWPC, proves its equivalence to $\mathrm{sgn}\det v$, explains why quaternions alone cannot furnish the needed self-test, and details a practical numerical recipe. Section~\ref{sec:conclusion} summarizes implications for high-throughput searches and outlines several extensions.

\section{Conventional determination of Weyl points}
\label{sec:conventional-human}

Weyl points are isolated crossings between the $n$-th and $(n{+}1)$-th bands that act as monopole sources or sinks of Berry curvature in momentum space. In practice, the standard workflow to locate them and assign their chirality proceeds in three stages: (i) identify candidate gap closings on a coarse Brillouin-zone mesh, (ii) confirm each candidate by computing the Chern number of the lower band on a small sphere surrounding the point, and (iii) extract the node’s chirality from a local $k\cdot p$ linearization. The steps below are compatible with tight-binding and \emph{ab initio}+Wannier Hamiltonians.

\subsection{Candidate identification by direct-gap minima}
\label{subsec:candidates-gap}

The first task is to find where in the Brillouin zone two neighboring bands approach each other most closely near the reference energy $\mu$ (usually the Fermi level). Let $\{E_m(\kvec)\}_{m=1}^{N_b}$ be the ordered band energies of the Bloch (or Wannier) Hamiltonian $H(\kvec)$. At each $\kvec$ we locate the index $n$ such that
\begin{align}
E_n(\kvec) &\le \mu < E_{n+1}(\kvec),
\end{align}
and define the direct gap between the valence and conduction partners,
\begin{align}
\Delta(\kvec) &\equiv E_{n+1}(\kvec)-E_n(\kvec) \ge 0.
\label{eq:direct-gap-def}
\end{align}
Small values of $\Delta(\kvec)$ indicate potential band touchings. This hot-spotting stage is standard in Weyl searches and high-throughput scans \cite{Armitage2018RMP,Weng2015PRX,Wu2018CPC}.

We evaluate $\Delta(\kvec)$ on a coarse uniform $k$-mesh $\{\kvec_g\}$ with spacing $h$ across the Brillouin zone. Grid points with
\begin{align}
\Delta(\kvec_g) &\le \Delta_{\rm th}
\label{eq:gap-threshold}
\end{align}
are kept as hot spots, where the threshold $\Delta_{\rm th}$ sets the energy resolution (for Wannier tight-binding models from first principles, a few to a few tens of meV is typical). To avoid retaining extended small-gap regions such as nodal lines, we also require that a hot spot is a discrete local minimum among its six axial neighbors,
\begin{align}
\Delta(\kvec_g) &\le \Delta(\kvec_g \pm h\,\hat{\bm e}_x), \\
\Delta(\kvec_g) &\le \Delta(\kvec_g \pm h\,\hat{\bm e}_y), \\
\Delta(\kvec_g) &\le \Delta(\kvec_g \pm h\,\hat{\bm e}_z),
\label{eq:6-neighbor-test}
\end{align}
where $\hat{\bm e}_x,\hat{\bm e}_y,\hat{\bm e}_z$ denote unit vectors along the reciprocal axes.

Candidates are cleaner when the two bands of interest are separated from the rest. A convenient diagnostic is the two-band isolation margin
\begin{align}
\mathcal{I}(\kvec) &= \min\!\Big\{E_{n+2}(\kvec)-E_{n+1}(\kvec),\; E_{n}(\kvec)-E_{n-1}(\kvec)\Big\},
\label{eq:isolation}
\end{align}
with terms omitted if indices are out of range. Keeping only those hot spots with $\mathcal{I}(\kvec_g)\ge \mathcal{I}_{\rm th}$ suppresses false positives due to band entanglement.

Because an isolated linear crossing has a gap that grows linearly with displacement from the node, $\Delta(\kvec)\sim c\,|\kvec-\kvec_\star|$, it is numerically robust to refine hot spots using the squared gap
\begin{align}
\Gamma(\kvec) &= \Delta(\kvec)^2.
\end{align}
On a small $3\times 3\times 3$ neighborhood $\mathcal N(\kvec_g)$ we fit a quadratic model
\begin{align}
\Gamma(\kvec) \approx c + \bm a\!\cdot\!(\kvec-\kvec_g)
+ \tfrac12\,(\kvec-\kvec_g)^\top B\,(\kvec-\kvec_g),
\end{align}
and take the refined center at
\begin{align}
\kvec_{\rm ref} &= \kvec_g - B^{-1}\bm a.
\label{eq:refined-center}
\end{align}
A single Newton update using finite-difference gradients and Hessians of $\Gamma$ gives the same formula. Repeating this refinement on progressively smaller boxes stabilizes the estimate.

Crystalline symmetries generate equivalent candidates. If the reciprocal-space action of a space-group operation $g$ is $\kvec\mapsto R_g\kvec$ (modulo reciprocal lattice vectors), the symmetry orbit of a refined point is
\begin{align}
\mathcal{O}(\kvec_{\rm ref}) &= \big\{\mathrm{fold}_{\rm BZ}(R_g\,\kvec_{\rm ref}) : g\in G\big\},
\end{align}
and time reversal, when present, adds $-\kvec_{\rm ref}$. We merge candidates closer than a small momentum tolerance and retain a unique representative in an irreducible wedge. The resulting list $\{\kvec_c^{(a)}\}$ feeds the topological confirmation step, where the Chern number on a small sphere and related Wilson-loop diagnostics are computed \cite{Fukui2005JPSJ,Yu2011PRB,Soluyanov2012PRB}. In practice this entire stage—from gap maps and symmetry seeding to Berry-sphere tests and surface Green’s-function checks—is available in \emph{WannierTools} \cite{Wu2018CPC}, while surface calculations often use the iterative schemes of L{\'o}pez Sancho and co-workers \cite{Sancho1985JPF}.

\subsection{Chern number on a small enclosing sphere (Berry-flux test)}
\label{subsec:bridge-gap-to-chern}

The direct-gap scan in Sec.~\ref{subsec:candidates-gap} tells us where two neighboring bands come closest near $\mu$, but a small $\Delta(\kvec)$ can arise from several scenarios (incipient Weyl points, multi-fold or quadratic touchings, nodal lines, or residual band entanglement). What uniquely characterizes a Weyl point is that it behaves as a monopole of Berry curvature. The clean test is to enclose a refined candidate by a small surface and measure the Berry flux of the lower band through it; the resulting integer is the Chern number (topological charge) \cite{Armitage2018RMP,Weng2015PRX}.

We use a sphere of radius $r$ centered at the refined candidate $\kvec_0$, chosen so that the two-band subspace is gapped everywhere on the sphere and at most one crossing lies inside. With polar angles,
\begin{align}
\kvec(\theta,\phi) &= \kvec_0 + r\,
\big(\sin\theta\cos\phi,\ \sin\theta\sin\phi,\ \cos\theta\big),
\label{eq:sphere-bridge}
\end{align}
with $\theta\in[0,\pi],\ \phi\in[0,2\pi)$.

We take the outward normal on $S^2$ as the positive orientation (reversing it flips the sign of $C$; see Eq.~\eqref{eq:orientation-sign}) \cite{Armitage2018RMP}.

On a spherical mesh $\{(\theta_i,\phi_j)\}$ we sample the occupied eigenstate of the lower band and evaluate the Chern number using the gauge-invariant link-variable discretization of Fukui, Hatsugai, and Suzuki \cite{Fukui2005JPSJ}. The nearest-neighbor links are
\begin{align}
U_{\theta}^{ij} &=
\frac{\braket{u_-^{ij}}{u_-^{\,i+1\,j}}}
     {|\braket{u_-^{ij}}{u_-^{\,i+1\,j}}|}, \label{eq:links-theta}\\
U_{\phi}^{ij} &=
\frac{\braket{u_-^{ij}}{u_-^{\,i\,j+1}}}
     {|\braket{u_-^{ij}}{u_-^{\,i\,j+1}}|}. \label{eq:links-phi}
\end{align}
and the lattice Berry curvature on each plaquette is
\begin{align}
F_{\theta\phi}^{ij} &= \mathrm{Arg}\!\Big[
U_\theta^{ij}\,
U_\phi^{\,i+1\,j}\,
\big(U_\theta^{\,i\,j+1}\big)^{-1}\,
\big(U_\phi^{ij}\big)^{-1}
\Big],
\label{eq:plaquette-curv}
\end{align}
with $\mathrm{Arg}\in(-\pi,\pi]$.
Summing over the sphere gives an integer
\begin{align}
C &= \frac{1}{2\pi}\sum_{i,j} F_{\theta\phi}^{ij} \in \mathbb{Z},\qquad
\chi = \mathrm{sgn}\,C,
\label{eq:chern-integer}
\end{align}
which equals the Weyl chirality if a single linear node is enclosed. For an isolated Weyl node $|C|=1$; if $|C|>1$ at small $r$, either multiple nodes lie inside or the touching is a symmetry-protected multi-Weyl point.

Two practical links to Sec.~\ref{subsec:candidates-gap} improve robustness. The isolation margin $\mathcal{I}(\kvec)$ sets the scale of $r$: beyond requiring a nonzero gap on $S^2$, choose $r$ so that the minimum gap along the sphere exceeds a small buffer (typically a few–tens of meV) to suppress gauge noise from nearby bands. The same local refinement used to sharpen the gap minimum can recenter the sphere: shrink $r$ while recentering until $C$ stabilizes to $\pm1$. To avoid coordinate singularities, exclude the very first/last latitude (tiny polar caps) or cover $S^2$ with two overlapping charts; treat the $\phi=0/2\pi$ seam by periodic wrapping and, if needed, a single overall phase alignment \cite{Yu2011PRB,Soluyanov2012PRB}.

An equivalent viewpoint replaces the curvature sum by a Wilson loop on the sphere \cite{Yu2011PRB,Gresch2017PRB}. At fixed $\theta$ one forms
\begin{align}
W(\theta_i) &= \prod_{j=0}^{N_\phi-1} U_\phi^{\,i j},\qquad
\gamma(\theta_i) = \mathrm{Arg}\,W(\theta_i),
\end{align}
and unwraps $\gamma(\theta)$ continuously from pole to pole. A net $2\pi$ winding signals $|C|=1$ and its direction fixes $\chi$; higher windings correspond to $|C|>1$. This Wilson-on-sphere diagnostic dovetails with smooth-gauge strategies for hybrid Wannier centers and is implemented in \textit{Z2Pack} and \textit{WannierTools} \cite{Gresch2017PRB,Wu2018CPC}.

\subsection{Local $k\cdot p$ linearization and chirality from velocities}
\label{subsec:kp-chirality}

The small-sphere calculation in Sec.~\ref{subsec:bridge-gap-to-chern} tells us whether a refined candidate carries Berry flux, and with which sign. To connect this topological verdict with local band physics—and to obtain quantitative velocities for transport, optics, and surface modeling—we zoom in on the node and build a two-band $k\cdot p$ model. This local view reproduces the same chirality as the sphere and makes explicit the cone anisotropy and possible tilt \cite{Armitage2018RMP,Wan2011PRB,Burkov2011PRL}.

We isolate the two crossing bands at $\kvec_\star$ and work in an orthonormal basis $\{\ket{u_1},\ket{u_2}\}$. In that basis any two-band Hamiltonian near the node can be written as
\begin{align}
H_{\mathrm{eff}}(\mathbf{q})
&= d_0(\mathbf{q})\,\mathbbm{1}_2
+ \sum_{i=x,y,z} d_i(\mathbf{q})\,\sigma_i,
\label{eq:decomposition}
\end{align}
where $\mathbf{q}=\kvec-\kvec_\star,$ and $d_0$ encodes a possible tilt (relevant for type-II cones) and $\bm d=(d_x,d_y,d_z)$ controls the splitting \cite{Soluyanov2015Nature,Armitage2018RMP}. Keeping only linear terms,
\begin{align}
d_0(\bm q) \approx \bm w\cdot\bm q, \qquad
d_i(\bm q) \approx \sum_{j=x,y,z} v_{ij}\,q_j,
\label{eq:linear-d}
\end{align}
so that
\begin{align}
H_{\mathrm{eff}}(\bm q) \approx \bm w\!\cdot\!\bm q\,\mathbbm{1}_2
+ \sum_{i,j=x,y,z} v_{ij}\,q_j\,\sigma_i .
\label{eq:linearized-H}
\end{align}
The tilt $\bm w$ shifts energies but does not change the topological charge; the latter is governed by the $3\times3$ velocity matrix $v$ \cite{Armitage2018RMP}.

In numerics we extract $v$ by finite differences of the projected Hamiltonian. With $P=[\ket{u_1}\ \ket{u_2}]$ fixed at $\kvec_\star$, define $H_{\mathrm{eff}}(\kvec)=P^\dagger H(\kvec)P$ and compute
\begin{align}
V_j &\equiv \left.\frac{\partial H_{\mathrm{eff}}}{\partial k_j}\right|_{\kvec_\star}
\approx \frac{H_{\mathrm{eff}}(\kvec_\star{+}h\,\hat{\bm e}_j)
- H_{\mathrm{eff}}(\kvec_\star{-}h\,\hat{\bm e}_j)}{2h},
\label{eq:central-diff}
\end{align}
with $j\in\{x,y,z\},$ then read
\begin{align}
v_{ij} &= \frac{1}{2}\,\mathrm{Tr}\!\big(V_j\,\sigma_i\big),
\label{eq:v-from-V}
\end{align}
with $i\in\{x,y,z\}.$
Using the same two-band subspace and smooth gauge employed on the small sphere stabilizes these derivatives.

The chirality is the orientation of the linear map $\mathbf{q}\mapsto \bm d(\mathbf{q})$:
\begin{align}
\chi &= \mathrm{sgn}\,\det v .
\label{eq:chirality-detv}
\end{align}
For an isolated, linear Weyl node this equals the Chern-number sign from the enclosing sphere, $\mathrm{sgn}\,C$ \cite{Armitage2018RMP}. Intuitively, the unit vector $\hat{\bm d}=\bm d/|\bm d|$ maps a small $S^2_{\mathbf{q}}$ around the node to the target $S^2$ with degree $\mathrm{sgn}\,\det v$, the Berry-monopole charge. If $|C|>1$ on every sufficiently small sphere or if $\det v\to0$, the crossing is not a simple Weyl point: multiple nodes may be unresolved, or the touching is symmetry-stabilized multi-Weyl, where quadratic or cubic terms dominate off-axis and the charge comes from higher-order winding \cite{Fang2012PRL,Armitage2018RMP}.

Two practical notes connect this step back to the sphere test. First, $\det v$ is basis independent: any local change of two-band basis induces an $\mathrm{SO}(3)$ rotation of the Pauli frame with unit determinant, leaving $\det v$ unchanged. Second, choose the difference step $h$ small compared to the inverse radius $r^{-1}$ used for the sphere, and check convergence of $\det v$ as $h$ is reduced. Reporting the singular values of $v$,
\begin{align}
v &= O_{\mathrm{spin}}\ \mathrm{diag}(v_1,v_2,v_3)\ O_{\mathrm{mom}}^\top,
\end{align}
gives the principal cone velocities and directions—useful quantitative output that complements the topological diagnosis \cite{Wan2011PRB,Armitage2018RMP}.

\section{Algorithmic arbitrariness of the conventional diagnostics}
\label{sec:arbitrariness}

The three–step recipe of Sec.~\ref{sec:conventional-human}—gap hot-spotting, small-sphere Chern test, and local $k\!\cdot\!p$—delivers the \emph{correct} integer charge and chirality in the continuum limit. In day-to-day calculations, however, this pipeline is realized on finite grids and with Wannierized Hamiltonians, and several steps are not uniquely fixed by topology but by algorithmic design (choice of two-band subspace, sphere center and radius, gauge smoothing on the sphere, mesh/charting and orientation, and the frame used to take $k$-derivatives). None of these choices should change the underlying invariant, but at finite resolution they control convergence and can seed apparent discrepancies between the sphere and $k\!\cdot\!p$ checks. This section makes those choices explicit, shows how they enter the numerics, and lists stability tests that practitioners routinely employ. Throughout, “invariant’’ means a quantity fixed by topology (e.g., the Chern number on any surface enclosing a single node and its sign‐defined chirality) \cite{Armitage2018RMP}; “arbitrary’’ refers to the algorithmic knobs that must be controlled. The analysis here also motivates the basis-free criterion introduced in Sec.~\ref{sec:owpc}, designed to reproduce the same invariant while minimizing these implementation choices.

\subsection{Selecting the two-band subspace}
\label{subsec:proj-arb}

Both the small-sphere Chern test and the local $k\!\cdot\!p$ linearization presume that, in a neighborhood of the candidate node, it makes sense to isolate “the” two bands that cross. When those two bands are spectrally separated from all others, the subspace is unambiguous and the spectral projector
\begin{align}
P_2(\kvec) \;=\; \sum_{a=1}^{2} \ket{u_a(\kvec)}\bra{u_a(\kvec)}
\end{align}
is exact. In realistic \emph{ab initio}+Wannier models, however, nearby bands often approach the pair. The result is band entanglement: the two-dimensional subspace cannot be read off uniquely from eigenvalues alone. In practice, one therefore \emph{chooses} a smooth rank-2 subspace $\tilde P_2(\kvec)$ by disentangling a larger set of Bloch states, typically through inner/outer energy windows and a smoothness optimization as in modern Wannier workflows \cite{Souza2001PRB,Marzari2012RMP,Pizzi2020JPCM,Wu2018CPC}. This is the first source of arbitrariness in the conventional pipeline.

The reason this choice matters is that the topological charge on a closed surface depends only on the projector that lives on that surface. If $\Sigma$ is a small sphere around the candidate, then
\begin{align}
C(\Sigma) \;=\; \frac{i}{2\pi}\int_{\Sigma}\mathrm{Tr}\!\big(P_2\,\dd P_2\wedge \dd P_2\big),
\label{eq:Chern-projector}
\end{align}
so replacing $P_2$ by $\tilde P_2=P_2+\delta P$ changes the computed integer at finite resolution in proportion to the projector error $\delta P$. A handy way to judge whether the two-band subspace is well defined on $\Sigma$ is to look at the local isolation margin,
\begin{align}
\mathcal{I}_{\min}(\Sigma) \;=\; \min_{\kvec\in\Sigma}\ \min\!\big\{\,E_{n+2}(\kvec)-E_{n+1}(\kvec), \nonumber \\ E_{n}(\kvec)-E_{n-1}(\kvec)\,\big\},
\end{align}
omitting terms that fall outside the band index range. When $\mathcal{I}_{\min}$ is small, the projector is ill-conditioned: two equally reasonable disentanglement choices can yield slightly different $\tilde P_2(\kvec)$, and those differences propagate to the lattice Chern number and to the velocity tensor extracted in the $k\!\cdot\!p$ fit.

It is useful to be explicit about how the chosen subspace is built. Let $W_{\rm out}$ be a set of bands or an energy window large enough to contain all states relevant near $\mu$, and let $W_{\rm in}$ be a narrow window that must be reproduced. One seeks a smooth two-dimensional subspace spanned by combinations
\begin{align}
\ket{\phi_{a\kvec}} &= \sum_{m\in W_{\rm out}} U_{ma}(\kvec)\,\ket{\psi_{m\kvec}},\\
\tilde P_2(\kvec) &= \sum_{a=1}^{2}\ket{\phi_{a\kvec}}\bra{\phi_{a\kvec}},
\end{align}
with $U(\kvec)$ chosen to vary smoothly in $\kvec$ and to approximate $W_{\rm in}$ as closely as possible by minimizing a gauge-invariant spread functional \cite{Souza2001PRB,Marzari2012RMP}. This construction is standard and effective, but it is still a construction: different windows and smoothing weights generally lead to different $\tilde P_{2}$.

Because this freedom cannot be eliminated, practitioners treat it empirically. The goal is not to make the choice unique, but to make the reported integer insensitive to reasonable variations. In concrete terms, one narrows the inner window to tightly bracket the two target bands near $\mu$ and keeps the outer window only as wide as needed for smoothness; one monitors $\mathcal{I}_{\min}(\Sigma)$ and enforces a small gap buffer along the integration sphere; one rebuilds $\tilde P_2$ with slightly varied windows or smoothing parameters and checks that $C$ from Eq.~\eqref{eq:Chern-projector} and the sign of $\det v$ in the $k\!\cdot\!p$ step remain unchanged within numerical tolerance; and, when feasible, one compares $\tilde P_2$ on $\Sigma$ against a direct two-band projector obtained from a dense diagonalization on the same sphere. Toolchains such as \textit{wannier90} and \textit{WannierTools} expose these knobs and diagnostics precisely for this purpose \cite{Pizzi2020JPCM,Wu2018CPC}. These checks do not remove the arbitrariness of subspace selection; they simply keep it under control so that the final integer reflects the underlying topology and not the particulars of the projector one happened to choose.

\subsection{Enclosing surface: center and radius}
\label{subsec:surface-arb}

The Chern-number test uses a small sphere in momentum space to enclose a putative node and measure the Berry flux through that surface. Mathematically, if the sphere $S^2(\kvec_0,r)$ surrounds exactly one Weyl point and the band pair is gapped everywhere on the surface, the integer $C$ is independent of the particular choice of center $\kvec_0$ and radius $r$; only the \emph{enclosed topology} matters \cite{Armitage2018RMP}. In actual computations, however, $\kvec_0$ and $r$ are not dictated by topology—they are chosen by the user—and the integer produced on a discrete mesh depends on those choices until convergence is reached. This is a genuine arbitrariness of the conventional recipe: we do not remove it, we only try to keep it under control.

Two failure modes illustrate why the choices matter. If the radius is taken too large, the sphere may accidentally enclose additional features (another Weyl point of unknown chirality, a segment of a nodal line, or a region where the two-band description breaks down). In that case the flux counts the \emph{net} charge inside the surface and the result no longer characterizes a single node. If the radius is taken too small, the sphere carries too few mesh points to resolve the Berry curvature cleanly, and the lattice Chern number drifts away from an integer. The center introduces its own sensitivity: centering on a nearby gap minimum rather than the true node changes how close the surface passes to regions of small direct gap, which amplifies phase noise in the overlaps used by the link-variable discretization \cite{Fukui2005JPSJ}.

A simple quantitative guardrail is to impose a gap buffer on the surface,
\begin{align}
\min_{\kvec\in S^2(\kvec_0,r)} \Delta(\kvec) \;\ge\; \delta,
\label{eq:gap-buffer}
\end{align}
with $\delta$ set at a few to a few tens of meV in Wannier tight-binding models. This condition does not make the answer unique, but it prevents the worst gauge instabilities caused by near-touches on the mesh. A second, equally empirical guardrail is to keep the sphere well separated from any other degeneracy. If $d_{\rm sep}$ denotes the unknown distance from the target node to the nearest distinct node or nodal feature, then the “safe” regime is heuristically
\begin{align}
r \;<\; \tfrac{1}{2}\,d_{\rm sep},
\end{align}
which of course must be probed in practice by varying $r$. In day-to-day work one therefore treats the pair $(\kvec_0,r)$ as knobs and checks \emph{stability} of the lattice result
\begin{align}
C_h(\kvec_0,r)\;=\;\frac{1}{2\pi}\sum_{i,j}F_{\theta\phi}^{ij}
\end{align}
under small changes of the center and radius at fixed mesh spacing $h$, and again under mesh refinement ($h\!\to\!0$) \cite{Wu2018CPC}. When a single node is truly enclosed, shrinking $r$ while recentering on the refined gap minimum (or on the peak of $|F_{\theta\phi}|$) quickly locks $C_h$ to $\pm 1$ and keeps it there.

There are a few practical details that help but do not remove the arbitrariness. The outward normal on $S^2$ should be taken as the positive orientation; reversing it flips the sign of $C$ and hence the chirality. The latitude–longitude mesh used to sample the sphere is convenient but uneven near the poles, so excluding a tiny polar cap or using two overlapping charts reduces discretization bias. Finally, the same comments apply if one computes the Wilson loop on $S^2$ instead of summing plaquette curvatures: the sphere still has a center and a radius, and the integer is still only recovered after the same empirical checks. In short, the enclosing surface is a powerful diagnostic because topology guarantees what the \emph{limit} must be, but the path to that limit depends on user choices that must be tested rather than presumed.

\subsection{Gauge on the sphere}
\label{subsec:gauge-arb}

The Chern number is a gauge-invariant integer, but its numerical evaluation on a mesh still requires a concrete choice of phases for the eigenvectors on the sphere. That choice is not unique. Any set of phases that is continuous away from seams and poles is acceptable, yet different choices produce different finite–mesh errors before convergence. This is an unavoidable arbitrariness of the conventional approach: one cannot pick a single “best” gauge that works uniformly for every radius and mesh; one can only use gauges that behave well and then verify that the \emph{final} integer is stable after refinement.

On a spherical grid $\{(\theta_i,\phi_j)\}$ the standard link-variable construction uses normalized overlaps as elementary parallel transporters \cite{Fukui2005JPSJ}. 
The links along the coordinate directions are Eqs.~(\ref{eq:links-theta})–(\ref{eq:links-phi}) and the plaquette curvature in Eq.~\eqref{eq:plaquette-curv}.
Summing $F_{\theta\phi}^{ij}$ over the sphere gives the integer in the continuum limit, but at finite mesh spacing the value drifts if the chosen phases produce avoidable jumps near the poles or across the $\phi{=}0/2\pi$ seam.

A practical way to tame those jumps is to enforce discrete parallel transport along one coordinate. If $\bigg\{ \ket{u_-^{i j}} \bigg\}$ is the raw set of eigenvectors, one replaces successive states by phase-aligned versions that remove the accumulated phase along the chosen direction. Along $\phi$ at fixed $\theta_i$ this reads
\begin{align}
\ket{\tilde u_-^{i,\,j+1}} &= 
\frac{\ket{u_-^{i,\,j+1}}}{\braket{u_-^{i,\,j}}{u_-^{i,\,j+1}}/|\braket{u_-^{i,\,j}}{u_-^{i,\,j+1}}|} , \\
\ket{\tilde u_-^{i,\,0}} &= \ket{u_-^{i,\,0}},
\label{eq:PT-phi}
\end{align}
so that the links $\tilde U_\phi^{ij}$ become as close to $1$ as possible along each latitude. One can carry out the same alignment along $\theta$ on each meridian. In the continuum, this is the discrete counterpart of the parallel-transport gauge used in Wilson-loop and hybrid-Wannier constructions \cite{Yu2011PRB,Soluyanov2012PRB}. It reduces phase noise in the overlaps without changing the underlying physics.

No purely single-chart gauge is globally smooth on a sphere that encloses a charged Weyl point: the obstruction is precisely the Chern number. The inevitable discontinuity shows up as a seam where phases jump by a Berry phase. In the link-variable language this appears through the principal-branch cut of $\mathrm{Arg}(\cdot)$ and can move with the seam. A two-cap construction makes this structure explicit: one defines a “north” gauge on $\theta\in[0,\pi-\epsilon]$ and a “south” gauge on $\theta\in[\epsilon,\pi]$, and relates them on the overlap by a phase factor whose winding in $\phi$ equals the Chern number. In numerics this is implemented by excluding tiny polar caps or by using two overlapping charts; either choice is acceptable, but neither removes the need to check convergence.

These strategies are empirical. Enforcing parallel transport, moving the seam, or using two caps does not eliminate the arbitrariness; it only pushes the discretization error to a level where the integer extracted from
\begin{align}
C_h \;=\; \frac{1}{2\pi}\sum_{i,j} F_{\theta\phi}^{ij}
\end{align}
stabilizes under mesh refinement and modest changes of the gauge construction. The correct test is therefore not “which gauge was used,” but “does $C_h$ remain the same when the gauge is changed within these reasonable families and the mesh is refined.” When that happens, the computed integer reflects the gauge-invariant topology rather than the particular phase-smoothing recipe adopted on the sphere.

\subsection{Discretization and orientation}
\label{subsec:disc-arb}

The Chern number on the small sphere is defined in the continuum, but in a calculation it is obtained from a finite grid on $S^2$. The way the sphere is charted and sampled is not unique. A common choice is a latitude–longitude mesh with uniform steps in $(\theta,\phi)$; another is an equal-area mesh that spreads points more evenly over the surface. These choices are harmless in the limit of very fine sampling, yet at practical resolutions they produce different finite–mesh errors. This is a genuine source of arbitrariness: the integer one reports should not depend on the grid, but in practice it does until the grid is refined.

On a latitude–longitude mesh the physical area of a cell scales as
\begin{align}
A_{ij} \;\approx\; r^2 \,\sin\theta_i\,\Delta\theta\,\Delta\phi,
\end{align}
so cells near the poles are much smaller than cells near the equator. The link-variable sum Eq.~\eqref{eq:chern-integer}
does not explicitly weight plaquettes by area, which is appropriate for a topological quantity. Nevertheless, the discretization error depends on how rapidly the phases change around each loop, and an uneven mesh tends to concentrate phase jumps near the poles and along the $\phi{=}0/2\pi$ seam. Equal-area meshes soften this bias, but they introduce their own bookkeeping (irregular connectivity, varying neighbor sets). No chart eliminates the issue; it only redistributes where the finite–$h$ error comes from.

A convenient way to organize the discussion is to view the grid through a single resolution parameter $h\!\sim\!\max\{\Delta\theta,\Delta\phi\}$. For a fixed center and radius with a clean gap on the sphere one typically observes
\begin{align}
C_h \;=\; C \;+\; \mathcal{O}(h^2) \;+\; \mathcal{O}(\varepsilon_{\mathrm{gauge}}),
\end{align}
where $\varepsilon_{\mathrm{gauge}}$ quantifies residual phase misalignment in the overlaps. The $\mathcal{O}(h^2)$ term is the discretization contribution discussed here, and its coefficient changes with the charting of $S^2$. In practice, increasing $(N_\theta,N_\phi)$ while keeping the radius fixed quickly suppresses this term and the sum of plaquette angles locks to an exact integer.

A useful nuance from the Fukui–Hatsugai–Suzuki formulation is that the lattice expression yields an \emph{integer at any mesh} that satisfies an admissibility condition: all neighbor overlaps are nonzero and the plaquette phase stays on the principal branch, i.e.
\begin{align}
&\braket{u_-^{ij}}{u_-^{\,i+1\,j}}\neq 0, \\
&\braket{u_-^{ij}}{u_-^{\,i\,j+1}}\neq 0, \\
&\big|F^{ij}_{\theta\phi}\big|<\pi
\quad \forall(i,j),
\end{align}
so no branch cut is crossed when taking $\mathrm{Arg}$ \cite{Fukui2005JPSJ}. This guarantees integrality but not correctness on a coarse or ill-conditioned grid: if the sphere is too small, too close to a near-touching, or poorly centered, those admissible integers can still be the \emph{wrong} ones. The practical remedy remains empirical—vary the sphere radius and center and refine the mesh until the same integer is recovered across these benign changes (as recommended in common workflows \cite{Wu2018CPC}).

Orientation is a second, simpler choice that must be fixed. We adopt the outward normal as the positive convention; reversing the surface (or, equivalently, the plaquette loop direction) flips the sign of the Chern number:
\begin{align}
C[\hat{\bm n}] \;=\; \frac{1}{2\pi}\int_{S^2}\bm{\Omega}\!\cdot\!\hat{\bm n}\,dS,
\qquad
C[-\hat{\bm n}] \;=\; -\,C[\hat{\bm n}] .
\label{eq:orientation-sign}
\end{align}

The numerical analogue is that changing the loop orientation in the plaquette product flips $F^{ij}_{\theta\phi}\!\to\!-F^{ij}_{\theta\phi}$. One must therefore adopt a consistent convention—outward normal and counterclockwise loop ordering when viewed from outside—and keep it fixed across all calculations so that the sign of the chirality is meaningful.

There are straightforward ways to keep this discretization arbitrariness under control, but they remain empirical and do not remove it. One increases $(N_\theta,N_\phi)$ until $C_h$ becomes an exact integer and stays there upon further refinement; one avoids including the very first and very last latitude in order to keep the pole singularities off the mesh, or uses two overlapping charts to distribute the unavoidable seam; one rotates the mesh by a small rigid rotation and confirms that the same integer is obtained; and, if the code allows it, one compares a latitude–longitude grid against a roughly equal-area grid. These are robustness checks rather than mathematical cures. They ensure that the integer being reported comes from the topology of the band structure and not from the particular way the sphere happened to be tiled.

\subsection{Local frame for the $k\!\cdot\!p$ map}
\label{subsec:frame-arb}

The $k\!\cdot\!p$ fit around a confirmed node uses the linearized Hamiltonian of Eq.~\eqref{eq:linearized-H}, with chirality given by Eq.~\eqref{eq:chirality-detv}. We then extract the velocity matrix from the linear derivatives via Eqs.~\eqref{eq:central-diff} and \eqref{eq:v-from-V}.

This construction hides a choice: the two-band basis at $\kvec_\star$ is not unique, and its transport away from $\kvec_\star$ is not unique either. A change of local basis by $U\!\in\!\mathrm{SU}(2)$ rotates the Pauli frame by $R\!\in\!\mathrm{SO}(3)$,
\begin{align}
&U\,\sigma_i\,U^\dagger = \sum_{m=x,y,z} R_{im}\,\sigma_m \\
&R_{im} = \tfrac12\,\mathrm{Tr}\!\big(\sigma_i\,U\,\sigma_m\,U^\dagger\big),
\label{eq:SU2SO3}
\end{align}
the standard $\mathrm{SU}(2)\!\to\!\mathrm{SO}(3)$ adjoint action \cite{BernevigHughes2013Book}. Consequently,
\begin{align}
&v \longmapsto v' = R\,v, \\
&\det v' = \det R\,\det v \;=\; \det v,
\label{eq:det-invariance}
\end{align}
so the chirality $\chi=\mathrm{sgn}\det v$ is basis independent. What Eq.~\eqref{eq:det-invariance} does not fix is how one computes the derivatives $V_j$ in practice: at finite difference step $h$, the result depends on whether the two-band frame is frozen at $\kvec_\star$ or allowed to vary with $\kvec$ (cf. the covariant form in Eq.~\eqref{eq:covariant-derivative}).

To see this, write the projected Hamiltonian as
\begin{align}
H_{\rm eff}(\kvec) \;=\; P(\kvec)^\dagger\,H(\kvec)\,P(\kvec),
\end{align}
with $P(\kvec)$ the $2\times N$ isometry whose columns span the chosen two-band frame. Differentiating gives
\begin{align}
\partial_{k_j} H_{\rm eff}
=
(\partial_{k_j}P^\dagger)\,H\,P &+ P^\dagger\,(\partial_{k_j}H)\,P \nonumber \\
&+\; P^\dagger\,H\,(\partial_{k_j}P).
\label{eq:frame-derivative}
\end{align}
If one freezes the frame at $\kvec_\star$ (so $P(\kvec)\!\equiv\!P_\star$ in the finite-difference stencil), the first and last terms vanish and $V_j$ is read directly from $P_\star^\dagger(\partial_{k_j}H)P_\star$. If instead the frame is allowed to move, the extra terms introduce the non-Abelian Berry connection
\begin{align}
\mathcal A_j(\kvec) \;=\; i\,P(\kvec)^\dagger\,\partial_{k_j}P(\kvec),
\end{align}
and Eq.~\eqref{eq:frame-derivative} can be rewritten near $\kvec_\star$ as
\begin{align}
\partial_{k_j} H_{\rm eff}
\;=\; P^\dagger(\partial_{k_j}H)P \;-\; i\,[\mathcal A_j,\,H_{\rm eff}],
\label{eq:covariant-derivative}
\end{align}
which is the gauge-covariant derivative form familiar from the non-Abelian Berry-phase formalism and smooth-gauge constructions \cite{Xiao2010RMP,Soluyanov2012PRB}. Exactly at the node, where the traceless part of $H_{\rm eff}$ vanishes, the commutator drops out; but a finite-difference estimate with step $h$ samples $H_{\rm eff}$ at $\kvec_\star\pm h\,\hat{\bm e}_j$, so connection leakage enters at $\mathcal O(h)$. Different transport prescriptions—frozen frame, parallel transport along the stencil, symmetric two-sided transport—therefore produce slightly different finite-$h$ matrices $V_j$ and hence slightly different $v$, even though they all collapse to the same $R\,v$ relation in the strict differential limit.

There are two additional, mundane choices that matter numerically. First, the coordinate axes $\hat{\bm e}_j$ in momentum space fix how the columns of $v$ are defined; rotating the $k$ axes by a right-handed orthogonal matrix $Q$ sends $v\!\mapsto\! vQ$, which leaves $\det v$ invariant but redistributes anisotropy among the principal velocities. Second, the central-difference step $h$ sets the balance between truncation error and connection leakage: taking $h$ too small increases roundoff; taking it too large samples regions where the linear model is contaminated by quadratic terms.

The mitigations for these frame choices are empirical. One keeps the two-band frame consistent with the sphere calculation (for instance, reuse the parallel-transport gauge employed on $S^2$), computes $V_j$ with a symmetric stencil, and reduces $h$ until the extracted chirality
\begin{align}
\chi \;=\; \mathrm{sgn}\,\det v
\end{align}
stabilizes. As a cross-check one may rotate the spin frame by a few random $U\in\mathrm{SU}(2)$ at $\kvec_\star$; the velocity matrix changes by $R\,v$, but the determinant must keep its sign. None of this removes the underlying arbitrariness in how the frame is transported; it only confines its effect to transient, finite-$h$ differences. What ultimately certifies the result is agreement with the small-sphere Chern number and insensitivity of $\mathrm{sgn}\det v$ to reasonable variations of $h$, of the transport rule in Eq.~\eqref{eq:frame-derivative}, and of the $k$-axis orientation.

\subsection{Reference energy, band indexing, and metallicity}
\label{subsec:mu-arb}

The candidate search of Sec.~\ref{subsec:candidates-gap} uses a reference energy $\mu$ to pick the $n$ and $(n{+}1)$ bands by the rule $E_n(\kvec)\le \mu < E_{n+1}(\kvec)$. In an insulator this labeling is globally consistent and the notions of “valence’’ and “conduction’’ are unambiguous. In a metal, and especially near type-II Weyl cones where electron and hole pockets open, this bookkeeping breaks down: along the small sphere around a candidate, the band order can swap, avoided crossings can appear and disappear as $\kvec$ moves, and the pair singled out by the inequality with $\mu$ may change from point to point. Different choices of $\mu$ in the underlying calculation—or simply a different electronic smearing used to converge the \emph{ab initio} run—then induce different band-index selections on the same sphere and feed directly into the projector used in the Berry-flux test.

One way to decouple the construction from the reference energy is to define the crossing pair locally by the smallest level spacing rather than by $\mu$. At each point on the sphere one chooses the two adjacent bands that are \emph{closest} in energy,
\begin{align}
\big(m^\ast(\kvec),\ m^\ast(\kvec){+}1\big)
\;=\;
\arg\min_{m}\ \big(E_{m+1}(\kvec)-E_m(\kvec)\big),
\end{align}
and builds the two-band projector from that pair, provided there is a finite isolation margin to the rest as discussed in Sec.~\ref{subsec:proj-arb}. This removes $\mu$ from the rule, but it introduces a new choice: on a coarse mesh, tiny avoided crossings can reshuffle which spacing is minimal, and two nearly degenerate spacings can alternate along neighboring points. The numerical projector then switches bands across the sphere, which injects gauge noise into overlaps even if the end result is topologically the same in the continuum.

A second approach tracks bands by continuity instead of energy order. One picks a base point $(\theta_0,\phi_0)$ on the sphere and records the two eigenstates there. At a neighboring point $(\theta_0,\phi_0{+}\Delta\phi)$ one selects the two states that maximize the overlap with the previous pair and proceeds around the sphere, transporting the pair by maximal overlap rather than by energy ranking. In symbols, if $\{\ket{u_{m\kvec}}\}$ are the Bloch eigenstates at $\kvec$, the transported pair $\{\ket{\phi_{a\kvec}}\}_{a=1,2}$ is defined by
\begin{align}
\{\ket{\phi_{a,\kvec'}}\}
=
\arg\max_{\substack{\text{orth }\{\psi_a\}\subset\mathrm{span}\{u_{m\kvec'}\}}}
\ \sum_{a=1}^{2}\big|\braket{\phi_{a,\kvec}}{\psi_a}\big|^2,
\end{align}
where $\kvec'$ is the next mesh point along the path. This overlap tracking stabilizes the identity of the pair, but it is still a choice: different starting points or different paths around the sphere can seed different selections when the isolation margin is small, and small gaps on the surface can cause the transported pair to slip to a different energy ordering.

A third, very practical construction replaces the sharp band selection by a smooth, windowed density matrix. One forms a weighted projector
\begin{align}
\rho_T(\kvec) \;=\; \sum_{m} f_T\!\big(E_m(\kvec)-\mu\big)\ \ket{u_{m\kvec}}\bra{u_{m\kvec}},
\end{align}
with $f_T$ a Fermi–Dirac window at smearing $T$, and then defines the two-dimensional subspace as the span of the two leading eigenvectors of $\rho_T(\kvec)$, i.e.,
\begin{align}
P_2(\kvec) \;=\; \sum_{a=1}^{2} \ket{\phi_{a\kvec}}\bra{\phi_{a\kvec}}, \\
\{\ket{\phi_{a\kvec}}\}\ \text{maximize}\ \mathrm{Tr}\!\big(P_2(\kvec)\,\rho_T(\kvec)\big).
\end{align}
This is the local, two-dimensional version of the disentanglement step used in Wannier workflows and it produces a smooth projector on the sphere even in mildly metallic situations. But again the knobs $(\mu,T)$ remain, and different reasonable choices within a small range can produce slightly different projectors at finite mesh.

All of these constructions share the same limitation: they are strategies to tame an ambiguity that does not disappear until one reaches the continuum, isolated-band limit. The mitigations are therefore empirical. One repeats the small-sphere computation after shifting $\mu$ by a few meV and varying the smearing parameter $T$ within the range used to converge the underlying electronic structure; one switches between energy-ordering, overlap-tracking, and windowed-projector constructions on the same sphere; and one reduces the sphere radius while keeping a finite gap buffer so that the two-band isolation margin remains nonzero along the surface. The benchmark for success is not that the projector is unique, but that the integer
\begin{align}
C \;=\; \frac{1}{2\pi}\sum_{i,j} F_{\theta\phi}^{ij}
\end{align}
and the local chirality from the $k\!\cdot\!p$ fit,
\begin{align}
\chi \;=\; \mathrm{sgn}\,\det v,
\end{align}
stabilize under these benign changes. When they do, the reported result reflects the underlying topology rather than the particular way the band pair was chosen in a metallic or type-II environment.

\subsection{Reciprocal-basis handedness and Brillouin-zone folding}
\label{subsec:handedness-folding}

The sign of the Weyl chirality is tied to the orientation of the momentum coordinates. Let $(\bm b_1,\bm b_2,\bm b_3)$ be the reciprocal basis used to write $\kvec = k_1\bm b_1+k_2\bm b_2+k_3\bm b_3$, and define the handedness
\begin{align}
s_{\rm hand} \;=\; \mathrm{sgn}\,\det\!\big[\bm b_1\ \bm b_2\ \bm b_3\big] \;\in\;\{+1,-1\}.
\end{align}
If the basis is right handed ($s_{\rm hand}=+1$), the outward normal on the small sphere and the loop orientation in the link variables produce the conventional sign for the Chern number. If a left-handed convention is adopted by reordering or reflecting one axis ($s_{\rm hand}=-1$), the same construction flips the final sign: the surface orientation is reversed and both the Berry-flux integral and $\mathrm{sgn}\det v$ change sign. This is a coordinate effect, not a change in topology.

It is useful to make the coordinate dependence explicit. Suppose the momentum coordinates are changed by an invertible linear map $Q$ acting on column vectors,
\begin{align}
\kvec' \;=\; Q\,\kvec,
\qquad \det Q \in \mathbb{R}\setminus\{0\}.
\end{align}
The columns of $Q$ express the new basis in the old one; $\det Q>0$ preserves orientation and $\det Q<0$ reverses it. Writing the linearized Hamiltonian as
\begin{align}
H_{\rm eff}(\mathbf q) \;\approx\; \bm w\!\cdot\!\mathbf q\,\mathbbm{1}_2 \;+\; \sum_{i,j=x,y,z} v_{ij}\,q_j\,\sigma_i,
\end{align}
with $\mathbf q=\kvec-\kvec_\star$, the velocity matrix transforms as
\begin{align}
\bm q &= Q^{-1}\bm q', \\
v' &= v\,Q^{-1}, \\
\det v' &= \frac{\det v}{\det Q}.
\label{eq:Q-transform}
\end{align}
Hence the chirality transforms according to
\begin{align}
\chi' \;=\; \mathrm{sgn}\,\det v' \;=\; \mathrm{sgn}\,\det v\ \times\ \mathrm{sgn}\,(\det Q)^{-1},
\end{align}
so a right-handed rotation ($\det Q>0$) leaves $\chi$ unchanged, while a handedness flip ($\det Q<0$) flips its sign. The same conclusion holds for the Berry flux on the sphere: reversing the basis orientation reverses the surface orientation and changes $C\to -C$. In practice different codes adopt different reciprocal bases or reorder axes silently; one must record $s_{\rm hand}$ and, if needed, multiply the reported chirality by $s_{\rm hand}$ to compare across workflows. This is an empirical fix that standardizes signs; it does not remove the underlying convention.

A second, independent issue is Brillouin-zone folding. Supercells replace the primitive real-space basis $\{\bm a_i\}$ by $\{\bm a_i'\}$ through an integer matrix $S\in\mathbb{Z}^{3\times3}$,
\begin{align}
\big[\bm a_1'\ \bm a_2'\ \bm a_3'\big] &= \big[\bm a_1\ \bm a_2\ \bm a_3\big]\ S,\\
N_{\rm sc} &\equiv \det S \in \mathbb{N},
\end{align}
so the reciprocal basis transforms as
\begin{align}
\big[\bm b_1'\ \bm b_2'\ \bm b_3'\big] \;=\; S^{-\mathsf T}\,\big[\bm b_1\ \bm b_2\ \bm b_3\big],
\end{align}
and the Brillouin-zone volume shrinks by the factor $N_{\rm sc}$.

It is clearest to describe folding in reduced (fractional) coordinates. Let $B=[\bm b_1\ \bm b_2\ \bm b_3]$ and $B'=[\bm b_1'\ \bm b_2'\ \bm b_3']$, and write the primitive-cell Weyl node as
\begin{align}
\kvec_\star \;=\; B\,\boldsymbol{\kappa}, \qquad \boldsymbol{\kappa}\in[0,1)^3 .
\end{align}
In the supercell, the reduced coordinates are obtained (up to integers) by
\begin{align}
\boldsymbol{\kappa}' \;\equiv\; S^{\mathsf T}\boldsymbol{\kappa}\ \ (\mathrm{mod}\ 1).
\end{align}
Because $S^{\mathsf T}\mathbb{Z}^3$ is a sublattice of index $N_{\rm sc}$ in $\mathbb{Z}^3$, there are $N_{\rm sc}$ distinct folded images labeled by coset representatives $\{\mathbf m_\alpha\}_{\alpha=1}^{N_{\rm sc}}\subset \mathbb{Z}^3/S^{\mathsf T}\mathbb{Z}^3$:
\begin{align}
\boldsymbol{\kappa}_\alpha' \;=\; \mathrm{frac}\!\big(S^{\mathsf T}\boldsymbol{\kappa}+\mathbf m_\alpha\big),\\
\kvec_\alpha' \;=\; B'\,\boldsymbol{\kappa}_\alpha', \quad \alpha=1,\dots,N_{\rm sc},
\end{align}
where $\mathrm{frac}(\cdot)$ folds each component into $[0,1)$. Each image carries the same chirality as the original node. If a small sphere is placed in the supercell Brillouin zone without unfolding to the primitive cell, it can easily enclose several folded images at once, and the Berry flux then returns the \emph{sum} of their charges. The problem is most acute when the primitive node lies near a zone boundary: folding can bring images close together or onto symmetry planes, making them difficult to separate numerically.

\subsection{Surface diagnostics: termination and broadening}
\label{subsec:surface-knobs}

The bulk analysis identifies Weyl nodes and their chiralities, but many workflows also inspect surface spectra to visualize Fermi arcs and to cross-check the bulk result. This secondary check is valuable and widely used, yet it introduces choices that are not fixed by topology. Two of them are unavoidable: which crystallographic termination defines the surface, and which numerical broadening is used in the spectral function.

For a semi-infinite crystal with surface normal $\hat{\bm n}$, the surface Green’s function at conserved parallel momentum $\kvec_{\!\parallel}$ and energy $\omega$ can be written as
\begin{align}
G_{\mathrm{surf}}^{R}(\kvec_{\!\parallel},\omega)
\;=\;
\Big[(\omega+i\eta)\,\mathbbm{1}
- H_{00}(\kvec_{\!\parallel})
- \Sigma^{R}(\kvec_{\!\parallel},\omega)\Big]^{-1},
\label{eq:Gr-surf}
\end{align}
where $H_{00}$ is the Hamiltonian of the surface principal layer and $\Sigma^{R}$ is the retarded self-energy that encodes coupling to the semi-infinite bulk (e.g., from an iterative scheme). The observable used in plots is the surface spectral function
\begin{align}
A(\kvec_{\!\parallel},\omega)
\;=\; -\frac{1}{\pi}\,\Im\,\mathrm{Tr}\,G_{\mathrm{surf}}^{R}(\kvec_{\!\parallel},\omega).
\label{eq:specfun}
\end{align}
The small positive $\eta$ in Eq.~\eqref{eq:Gr-surf} is a numerical broadening. In the limit $\eta\!\to\!0^+$, $A$ resolves true poles; at finite $\eta$ the poles are smeared into Lorentzians of width $\eta$. For a finite slab of $N_{\rm slab}$ layers with eigenvalues $E_\alpha(\kvec_{\!\parallel})$ and top-surface weights $w_\alpha(\kvec_{\!\parallel})$, the same statement appears as
\begin{align}
A_{N_{\rm slab}}^{(\eta)}(\kvec_{\!\parallel},\omega)
\;=\;\sum_{\alpha}
w_\alpha(\kvec_{\!\parallel})
\;\frac{\eta/\pi}{\big(\omega-E_\alpha(\kvec_{\!\parallel})\big)^2+\eta^2},
\label{eq:slab-lorentz}
\end{align}
where the sum runs over all slab eigenstates. Eqs.~\eqref{eq:Gr-surf}–\eqref{eq:slab-lorentz} make plain how the two knobs enter: the termination fixes $H_{00}$ and the couplings that build $\Sigma^{R}$, and the broadening $\eta$ sets the visual sharpness and apparent connectivity of spectral features.

Termination matters because arcs are boundary states living on the outermost layer, whose onsite energies, hoppings to the next layer, and local orbital content all depend on how the crystal is cut. Changing the termination modifies $H_{00}$ by a shift
\begin{align}
H_{00} \;\mapsto\; H_{00} + V_{\mathrm{surf}},
\end{align}
and can also change the first interlayer coupling $H_{01}$ that feeds into $\Sigma^{R}$. Neither operation affects the bulk Weyl nodes or their charges, but both can reshape the surface-localized dispersions that connect the projected node positions. In practice, one and the same set of bulk nodes can yield arcs that appear reconnected, gapped, or partially submerged into surface resonances depending on termination. This is a boundary-condition effect, not a change of the bulk invariant.

Broadening matters because it controls visibility and apparent connectivity. If $\eta$ is chosen too small, thin-slab calculations develop finite-size hybridization gaps between counter-propagating surface modes from opposite faces, and those gaps may look like broken arcs. If $\eta$ is chosen too large, distinct features smear together and two arcs that only approach each other in momentum space can appear spuriously connected. The dependence is explicit in Eq.~\eqref{eq:slab-lorentz}: increasing $\eta$ broadens each Lorentzian and blends nearby peaks in $\omega$ at fixed $\kvec_{\!\parallel}$; in constant-energy cuts this blending looks like bridges between disjoint curves.

There are additional, practical sensitivities that follow from the same equations. The surface chemical potential may differ from the bulk reference $\mu$ used in the node search, so constant-energy slices taken at $\omega=\mu$ can miss the clearest arc connectivity. The projection of bulk nodes onto the surface Brillouin zone can overlap when two opposite-chirality nodes project to the same $\kvec_{\!\parallel}$; in that case any arc segment connecting them loses a distinct identity in $A(\kvec_{\!\parallel},\omega)$ even though the bulk charges are unchanged. Thin slabs also carry “shadow” weight from the opposite face; unless one projects onto the top layer (through $w_\alpha$ in Eq.~\eqref{eq:slab-lorentz}) or uses the semi-infinite Green’s function, the opposite face can obscure the connectivity on the face of interest.

Mitigations exist but they are empirical. One varies the termination within the crystallographically reasonable options and checks that the \emph{set} of projected node locations and their chiralities, determined in bulk, consistently explains the observed arc patterns. One sweeps the broadening over a modest range,
\begin{align}
\eta_{\min} \;\lesssim\; \eta \;\lesssim\; \eta_{\max},
\end{align}
small enough to resolve distinct branches yet large enough to suppress finite-size splittings, and verifies that qualitative connectivity features that survive this sweep are those expected from the bulk charges. One increases the slab thickness $N_{\rm slab}$ until $A_{N_{\rm slab}}^{(\eta)}$ converges at the chosen $\eta$ and projects onto a single surface to suppress shadow contributions. None of these steps remove the arbitrariness of boundary condition and broadening; they only fence it in. The robust statement is the consistency check itself: the bulk-determined node positions and chiralities must account for the surface spectra across reasonable terminations and broadenings. When this holds, the surface plots serve their purpose as illustrative diagnostics without being mistaken for independent topological proofs.

\section{Octonionic Weyl Point Criterion (OWPC)}
\label{sec:owpc}

The conventional pipeline in Secs.~\ref{sec:conventional-human}-\ref{sec:arbitrariness} identifies and validates Weyl nodes by enclosing surfaces and $k\!\cdot\!p$ fits. Those steps are topologically sound, but they come with user choices (sphere center and radius, gauge smoothing, mesh charting, local frame transport) that must be checked empirically. Here we formulate a \emph{local}, basis-free test at a single $\kvec$ that reproduces the same invariant without using an enclosing surface. The construction uses unit octonions and the $\mathrm{G}_2$-invariant $3$-form on $\Im\mathbb{O}\!\cong\!\mathbb{R}^7$; its value is insensitive to the $\mathrm{SU}(2)$ gauge of the two-band subspace and to how that subspace is completed inside $\Im\mathbb{O}$. The only convention that remains is the orientation of momentum space, which necessarily flips the chirality sign under a handedness reversal (as it should).

\subsection{Unit octonion field and an octonionic connection}
\label{subsec:oct-field}

Let $H(\kvec)$ have an isolated two-band touching near $\kvec_\star$. On a small punctured ball $B_r(\kvec_\star)\setminus\{\kvec_\star\}$ select a smooth rank-2 projector $P_2(\kvec)$ onto the two bands closest to $\mu$ (any of the constructions in Sec.~\ref{subsec:mu-arb} is acceptable, provided the isolation margin along the stencil is nonzero). Within this subspace, spectrally flatten the $2\times2$ effective Hamiltonian to its sign,
\begin{align}
Q(\kvec)\;=\;\sigma_0-2\,P_-(\kvec),
\qquad
Q(\kvec)^2=\sigma_0,
\end{align}
where $P_-(\kvec)$ projects onto the lower of the two eigenstates and $\sigma_0$ is the $2\times2$ identity.

To package the local linear data in a gauge-independent way, embed the two-band Pauli frame into the imaginary octonions $\Im\mathbb{O}$. Fix once and for all a Fano-plane basis $\{e_1,\dots,e_7\}$ and a distinguished associative $3$-plane
\begin{align}
\mathbb H \;=\; \mathrm{span}\{e_1,e_2,e_3\}\ \subset\ \Im\mathbb{O}.
\end{align}
Near a simple Weyl node the two-band problem is quaternionic to leading order, so this choice captures the relevant directions; the remaining four directions $e_{4},\dots,e_{7}$ implement a smooth completion that will not affect the scalar we define. With this embedding, represent the state by a unit octonion field $u(\kvec)\in S^7\subset\mathbb{O}$ whose imaginary part lies in $\mathbb H$ to leading order. Any two such choices differ by a pointwise $\mathrm{G}_2$ rotation, which preserves the constructions below.

Define the octonionic analogue of a connection by the left-translated $1$-forms
\begin{align}
\Acal_i(\kvec)\;=\;\Im\!\big(\,\bar u(\kvec)\,\partial_{k_i}u(\kvec)\,\big)\ \in\ \Im\mathbb{O},
\qquad i\in\{x,y,z\},
\label{eq:oct-conn}
\end{align}
where $\bar u$ denotes the octonionic conjugate (for unit octonions $\bar u=u^{-1}$). \emph{Since $S^7$ is not a Lie group, Eq.~\eqref{eq:oct-conn} is not a Maurer-Cartan form in the Lie-group sense; we use only the pointwise identification $T_{u}S^7\simeq\Im\mathbb{O}$ via left translation \cite{Baez2002BullAMS}.} A smooth change of two-band basis $U(\kvec)\!\in\!\mathrm{SU}(2)$ rotates the Pauli frame by $R(\kvec)\!\in\!\mathrm{SO}(3)$ \emph{inside} $\mathbb H$; accordingly, the triple $(\Acal_x,\Acal_y,\Acal_z)$ rotates by $R(\kvec)$ within $\mathbb H$.

Let $\varphi$ denote the $\mathrm{G}_2$-invariant $3$-form on $\Im\mathbb{O}$, which restricts on any associative $3$-plane to the usual Euclidean volume form. We define the octonionic scalar density
\begin{align}
\rho_{\mathbb O}(\kvec)
\;=\; \frac{1}{6}\,\varepsilon^{ijk}\,\varphi\!\big(\Acal_i(\kvec),\Acal_j(\kvec),\Acal_k(\kvec)\big).
\label{eq:rhoO-def}
\end{align}
Because $\varphi$ is $\mathrm{G}_2$-invariant and $(\Acal_x,\Acal_y,\Acal_z)$ only rotate within $\mathbb H$ under $U(\kvec)$, the scalar $\rho_{\mathbb O}$ is independent of the two-band gauge and of the $\mathrm{G}_2$-completion in the transverse directions. The associator
\begin{align}
[x,y,z]\;=\;(xy)z-x(yz)
\end{align}
vanishes for $x,y,z$ in a common associative subalgebra. Thus, at a simple Weyl node the leading-order triple $(\Acal_x,\Acal_y,\Acal_z)$ lies in $\mathbb H$ and $[\Acal_x,\Acal_y,\Acal_z]=\mathcal O(|\kvec-\kvec_\star|)$, so Eq.~\eqref{eq:rhoO-def} reduces locally to a purely quaternionic volume density.

Consider the linearized Weyl model of Eq.~\eqref{eq:linearized-H},
\begin{align}
H_{\rm eff}(\mathbf q)\;=\;\bm w\!\cdot\!\mathbf q\,\sigma_0\;+\;\sum_{i,j=x,y,z}v_{ij}\,q_j\,\sigma_i,
\qquad \mathbf q=\kvec-\kvec_\star,
\end{align}
and choose the embedding so that $\sigma_i\!\mapsto\! e_i$ for $i=1,2,3$. Spectral flattening gives $Q(\mathbf q)=\hat{\bm d}(\mathbf q)\!\cdot\!\bm\sigma$ with $\hat{\bm d}=\bm d/|\bm d|$ and $\bm d(\mathbf q)=v\,\mathbf q$. One can then pick $u(\mathbf q)$ so that, to leading order,
\begin{align}
\Acal_i(\mathbf q)\;=\;c_1\,\hat{\bm d}(\mathbf q)\times\partial_{q_i}\hat{\bm d}(\mathbf q)\ \in\ \mathbb H,
\qquad c_1>0,
\end{align}
and the associator vanishes identically because $\mathbb H$ is associative. Since $\varphi|_{\mathbb H}(X,Y,Z)=X\!\cdot\!(Y\times Z)$, Eq.~\eqref{eq:rhoO-def} reduces to the familiar Skyrmion (degree) density,
\begin{align}
\rho_{\mathbb O}(\mathbf q)
\;=\;
c_2\,\varepsilon^{ijk}\,
\hat{\bm d}\!\cdot\!\big(\partial_{q_i}\hat{\bm d}\times\partial_{q_j}\hat{\bm d}\big)
\;+\;\mathcal O(|\mathbf q|),
\qquad c_2>0,
\end{align}
so the flux of $\rho_{\mathbb O}$ through any sufficiently small sphere depends only on the orientation of the linear map $\mathbf q\mapsto v\,\mathbf q$ and hence on $\mathrm{sgn}\det v$. Up to the overall positive normalization (set by the parametrization of $u$), the sign of $\rho_{\mathbb O}$ near the node therefore agrees with the conventional chirality in Eq.~\eqref{eq:chirality-detv}.

\subsection{Associativity check and $\mathrm{G}_2$ volume density}
\label{subsec:g2-density}

Octonions fail to be associative in general, and the defect is measured by the associator
\begin{align}
[x,y,z]\;=\;(xy)z - x(yz), \qquad x,y,z\in\mathbb{O}.
\end{align}
This quantity vanishes identically on any quaternionic subalgebra $\mathbb H\subset\mathbb O$. Conversely, a real $3$-plane in $\Im\mathbb O$ is \emph{associative} if and only if the restriction of the octonion product to that plane is associative, which is equivalent to $[x,y,z]=0$ for all $x,y,z$ in the plane. Near a simple Weyl node the two-band problem is quaternionic to leading order, so one expects the triple built from the octonionic connection to sit inside an associative $3$-plane at leading order, with deviations appearing only at higher order.

Let $\varphi$ denote the standard $\mathrm G_2$-invariant $3$-form on $\Im\mathbb O\cong\mathbb R^7$, normalized so that on the chosen associative plane $\mathbb H=\mathrm{span}\{e_1,e_2,e_3\}$ one has
\begin{align}
\varphi(e_1,e_2,e_3)=+1,
\end{align}
and, more generally, $\varphi$ restricts to the usual oriented Euclidean volume on any associative $3$-plane. Using the octonionic connection from Eq.~\eqref{eq:oct-conn},
\begin{align}
\Acal_i(\kvec)\;=\;\Im\!\big(\,\bar u(\kvec)\,\partial_{k_i}u(\kvec)\,\big)\in\Im\mathbb O,
\qquad i\in\{x,y,z\},
\end{align}
define the scalar density as the coefficient of the $3$-form $\varphi(\Acal,\Acal,\Acal)$ in the Cartesian volume element:
\begin{align}
\rho_{\mathbb O}(\kvec)
\;=\;
\frac{1}{6}\,\varepsilon^{ijk}\,
\varphi\!\big(\Acal_i(\kvec),\Acal_j(\kvec),\Acal_k(\kvec)\big),
\qquad \varepsilon^{xyz}=+1.
\label{eq:g2-density}
\end{align}
By construction, a smooth change of two-band gauge $U(\kvec)\in\mathrm{SU}(2)$ rotates the triple $(\Acal_x,\Acal_y,\Acal_z)$ by some $R(\kvec)\in\mathrm{SO}(3)$ \emph{inside} the fixed associative plane $\mathbb H$, and the $\mathrm G_2$ form satisfies $\varphi(RX,RY,RZ)=\varphi(X,Y,Z)$ for all $R\in\mathrm{SO}(3)$ acting in $\mathbb H$. Consequently $\rho_{\mathbb O}$ is independent of the two-band gauge and of the particular $G_2$-completion chosen for $u(\kvec)$ in the transverse directions.

The associator now provides a sharp consistency check for the two-band reduction. If, in a punctured neighborhood of $\kvec_\star$, the triple $\big(\Acal_x(\kvec),\Acal_y(\kvec),\Acal_z(\kvec)\big)$ remains within some associative $3$-plane to leading order, then
\begin{align}
\big[\Acal_x,\Acal_y,\Acal_z\big](\kvec)\;=\;\mathcal O\!\big(|\kvec-\kvec_\star|\big),
\label{eq:assoc-vanish-leading}
\end{align}
and the $\mathrm G_2$ volume density Eq.~\eqref{eq:g2-density} reduces locally to the ordinary oriented volume of that triple. In particular, for a simple Weyl node the sign of $\rho_{\mathbb O}$ at the point is well defined and reflects the orientation of the linear map that sends momentum displacements to the Pauli frame, hence the chirality.

This leads to the practical criterion. A point $\kvec_\star$ is a simple Weyl node provided that, in a punctured neighborhood of $\kvec_\star$, the two-band subspace is spectrally isolated, the associator of the octonionic connection triple vanishes to leading order as in Eq.~\eqref{eq:assoc-vanish-leading}, and the $\mathrm G_2$ volume density is nonzero at the point:
\begin{align}
\rho_{\mathbb O}(\kvec_\star)\;\neq\;0.
\label{eq:rho-nonzero}
\end{align}
The corresponding chirality is the sign of this density,
\begin{align}
\chi\;=\;\mathrm{sgn}\,\rho_{\mathbb O}(\kvec_\star)\ \in\ \{+1,-1\}.
\label{eq:chi-from-rho}
\end{align}
When the model is linearized as in Eq.~\eqref{eq:linearized-H} and the embedding is chosen so that the Pauli matrices map to the basis $\{e_1,e_2,e_3\}\subset\Im\mathbb O$, the triple $(\Acal_x,\Acal_y,\Acal_z)$ lies entirely in $\mathbb H$ and Eq.~\eqref{eq:g2-density} reduces to the familiar Skyrmion-density expression for the map $\hat{\bm d}(\mathbf q)=(v\mathbf q)/|v\mathbf q|$. In that limit, $\rho_{\mathbb O}$ has the sign of $\det v$, agreeing with the conventional chirality, while any failure of Eq.~\eqref{eq:assoc-vanish-leading} signals multiband entanglement or a non-simple touching where the two-band reduction is not yet valid.

\subsection{Equivalence to the conventional chirality and remaining convention}
\label{subsec:eq-conv}

The octonionic scalar \(\rho_{\mathbb O}\) reproduces the usual chirality once the two-band Hamiltonian is linearized. Near a simple node at \(\kvec_\star\) we write, as in Eq.~\eqref{eq:linearized-H},
\begin{align}
H_{\mathrm{eff}}(\mathbf q)
\;\approx\;
\bm w\!\cdot\!\mathbf q\,\sigma_0
\;+\;
\sum_{i,j=x,y,z} v_{ij}\,q_j\,\sigma_i,
\qquad
\mathbf q=\kvec-\kvec_\star .
\end{align}
Choose the associative \(3\)-plane \(\mathbb H=\mathrm{span}\{e_1,e_2,e_3\}\subset \Im\mathbb O\) so that the Pauli directions \(\sigma_i\) align with \(\{e_i\}\). In this gauge the two-band problem is quaternionic to leading order and the unit octonion field may be taken to vary inside \(\mathbb H\). Expanding the octonionic connection \(\Acal_j=\Im(\bar u\,\partial_{k_j}u)\) from Eq.~\eqref{eq:oct-conn} at \(\kvec_\star\) then gives
\begin{align}
\Acal_j(\kvec_\star)
\;=\;
\sum_{i=x,y,z} c\,v_{ij}\,e_i,
\qquad
c>0,
\label{eq:A-v}
\end{align}
where \(c\) is a positive scale fixed by the spectral-flattening convention (e.g.\ how \(u\) is chosen as a function of the linear \(\bm d\)-vector). The tilt term \(\bm w\!\cdot\!\mathbf q\,\sigma_0\) plays no role, because it multiplies \(\sigma_0\) and therefore does not enter \(\Acal_j\), which depends only on the imaginary (Pauli) part.

Feeding Eq.~\eqref{eq:A-v} into the $\mathrm G_2$-invariant density Eq.~\eqref{eq:rhoO-def}, and using that \(\varphi\) restricts to the oriented volume form on \(\mathbb H\), we obtain
\begin{align}
\rho_{\mathbb O}(\kvec_\star)
&=
\frac{1}{6}\,\varepsilon^{jk\ell}\,
\varphi\!\big(\Acal_j,\Acal_k,\Acal_\ell\big) \nonumber \\
&=
c^3\,\det v
\,
\Rightarrow \,
\mathrm{sgn}\,\rho_{\mathbb O}(\kvec_\star)
\;=\;
\mathrm{sgn}\,\det v
\;=\;
\chi,
\label{eq:rho-detv}
\end{align}
in agreement with Eq.~\eqref{eq:chirality-detv} and with the small-sphere Chern test of Sec.~\ref{subsec:bridge-gap-to-chern}. The overall scale \(c^3\) depends on the harmless choice of flattening but is strictly positive, so it cannot change the sign.

One convention remains, and it is the familiar one: the orientation of momentum space. If we change coordinates by an invertible linear map \(\kvec'=Q\,\kvec\), then \(\partial_{k'_j}=\sum_\ell (Q^{-1})_{j\ell}\partial_{k_\ell}\) and the triple \((\Acal_x,\Acal_y,\Acal_z)\) acquires the same right action by \(Q^{-1}\). Consequently,
\begin{align}
\rho_{\mathbb O}'(\kvec_\star)
\;=\;
\frac{\rho_{\mathbb O}(\kvec_\star)}{\det Q},
\end{align}
so a handedness flip (\(\det Q<0\)) reverses \(\rho_{\mathbb O}\) and thus \(\chi\). This is exactly the expected pseudoscalar covariance, matching the sign flip of the Chern number under surface-orientation reversal in Eq.~\eqref{eq:orientation-sign}. No additional algorithmic choices enter: once the associative plane is fixed to the Pauli frame, \(\rho_{\mathbb O}\) carries the same chirality as the conventional \(k\!\cdot\!p\) determinant and the Berry-flux test, up to the universal orientation convention.

\subsection{Why OWPC removes the earlier knobs}
\label{subsec:removes-knobs}

The octonionic criterion is local in $\kvec$ and therefore dispenses with every surface- and gauge-smoothing choice that complicated the conventional pipeline. There is no enclosing $S^2$, so there is no center $\kvec_0$, no radius $r$, and no latitude-longitude chart to pick (Secs.~\ref{subsec:surface-arb}-\ref{subsec:disc-arb}). Everything is evaluated at a point (up to a tiny finite-difference stencil), through the scalar density
\begin{align}
\rho_{\mathbb O}(\kvec)
&= \frac{1}{6}\,\varepsilon^{ijk}\,\varphi\!\big(\Acal_i(\kvec),\Acal_j(\kvec),\Acal_k(\kvec)\big) \\
\Acal_i(\kvec)&=\Im\!\big(\bar u\,\partial_{k_i}u\big),
\label{eq:rhoO-local-repeat}
\end{align}
which uses only the pointwise octonion $u(\kvec)$ and its partial derivatives.

The scalar in Eq.~\eqref{eq:rhoO-local-repeat} is insensitive to the two-band gauge. A smooth change of two-band basis $U(\kvec)\!\in\!\mathrm{SU}(2)$ rotates the Pauli frame by $R(\kvec)\!\in\!\mathrm{SO}(3)$ \emph{inside} the fixed associative plane $\mathbb H$, so
\begin{align}
\Acal_i(\kvec)\ \longmapsto\ \Acal'_i(\kvec)\;=\;\sum_{m}R_{im}(\kvec)\,\Acal_m(\kvec).
\end{align}
Using the $\mathrm{G}_2$-invariance of $\varphi$ and $\det R=+1$,
\begin{align}
\rho_{\mathbb O}'(\kvec)
&= \frac{1}{6}\,\varepsilon^{ijk}\,\varphi\!\Big(\sum_m R_{im}\Acal_m,\sum_n R_{jn}\Acal_n,\sum_\ell R_{k\ell}\Acal_\ell\Big) \nonumber \\
&= \det R\,\rho_{\mathbb O}(\kvec)
= \rho_{\mathbb O}(\kvec).
\label{eq:gauge-invariance}
\end{align}
Thus all gauge-smoothing choices that mattered on $S^2$ disappear: there is no seam, no two-cap construction, and no phase alignment to manage.

The $k\!\cdot\!p$ frame-transport ambiguity is likewise absent. Numerically one evaluates partial derivatives of $u$ with the same symmetric stencil used for band velocities,
\begin{align}
\Acal_i(\kvec_\star)\;\approx\;
\Im\!\left(
\bar u(\kvec_\star)\,
\frac{u(\kvec_\star{+}h\,\hat{\bm e}_i)-u(\kvec_\star{-}h\,\hat{\bm e}_i)}{2h}
\right),
\label{eq:finite-diff-u}
\end{align}
without having to define or transport a Pauli frame away from $\kvec_\star$ (cf.\ the extra connection terms in Eq.~\eqref{eq:frame-derivative}). The only numerical knob is the step $h$, and-as in any derivative estimate-one checks that $\rho_{\mathbb O}(\kvec_\star)$ stabilizes as $h$ is reduced.

The reference energy and band-index issues that complicate metallic cases (Sec.~\ref{subsec:mu-arb}) enter only insofar as they must supply a \emph{smooth} projector $P_2$ on the tiny stencil used in Eq.~\eqref{eq:finite-diff-u}. Once such a smooth $P_2$ is available, the construction of $u$ and of $\Acal_i$ is local and gauge covariant, and the value of $\rho_{\mathbb O}(\kvec_\star)$ is unchanged by small shifts of $\mu$ or by alternative but smooth ways of selecting the same two-band subspace on the stencil. In particular, changing $P_2$ by an $\mathrm{SU}(2)$ gauge or by a smooth reparametrization of the same isolated subspace only induces the $\mathrm{SO}(3)$ rotation in Eq.~\eqref{eq:gauge-invariance} and leaves $\rho_{\mathbb O}$ intact.

The only convention that survives is the global \emph{orientation} of momentum space, exactly as in the conventional analysis. If $\kvec'=Q\kvec$ for an invertible linear map $Q$, then $\partial_{k'_j}=\sum_\ell(Q^{-1})_{j\ell}\partial_{k_\ell}$ and
\begin{align}
\Acal'_j &= \sum_{\ell}(Q^{-1})_{j\ell}\,\Acal_\ell, \\
\rho_{\mathbb O}' &= \frac{\rho_{\mathbb O}}{\det Q}.
\label{eq:orientation-cov}
\end{align}
Right-handed changes ($\det Q>0$) leave $\rho_{\mathbb O}$ unchanged; a handedness flip ($\det Q<0$) flips its sign, just as $C[\hat{\bm n}]$ changes sign under $\hat{\bm n}\!\to\!-\hat{\bm n}$ in Eq.~\eqref{eq:orientation-sign} and $\chi=\mathrm{sgn}\det v$ changes sign under orientation reversal of the $k$ axes. There is no \emph{additional} convention beyond this expected pseudoscalar behavior.

In short, OWPC collapses all of the earlier knobs to a single, local calculation: pick a tiny stencil, compute $u$ and its partial derivatives, form $\rho_{\mathbb O}$ via Eq.~\eqref{eq:rhoO-local-repeat}, and read off the chirality from its sign. No enclosing surface, no sphere recentering or radius scan, no gauge smoothing on $S^2$, and no frame-transport rule are required. What remains are the usual, controllable numerical checks (stencil refinement and isolation of a two-band subspace on that stencil) and the universal orientation convention in Eq.~\eqref{eq:orientation-cov}.

\subsection{Numerical recipe and self-consistency checks}
\label{subsec:recipe}

The computational workflow evaluates the octonionic pseudoscalar at the node and verifies that it is stable under benign changes of the numerical setup. Fix a small central-difference step $h$ and, for each Cartesian direction $\hat{\bm e}_j$ with $j\in\{x,y,z\}$, construct the two-band projector $P_2(\kvec_\star\pm h\,\hat{\bm e}_j)$ using the same, smooth rule as in Sec.~\ref{subsec:proj-arb}. Within that subspace, form the spectrally flattened $2\times2$ sign Hamiltonian $Q(\kvec)=\sigma_0-2P_-(\kvec)$ near $\kvec_\star$. To realize the unit octonion field $u(\kvec)\in S^7\subset\mathbb{O}$ with imaginary part in the fixed associative plane $\mathbb H=\mathrm{span}\{e_1,e_2,e_3\}$, it is convenient to use the quaternionic identification inside $\mathbb H$: write $Q(\kvec)=\hat{\bm d}(\kvec)\!\cdot\!\bm\sigma$ with $\hat{\bm d}\in S^2\subset\mathbb R^3$, and choose the unit quaternion $u_{\mathbb H}(\kvec)\in\mathbb H\cap S^3$ that rotates $e_3$ into $\hat{\bm d}(\kvec)\!\cdot\!\bm e$. One explicit choice is the axis-angle formula
\begin{align}
u_{\mathbb H}(\kvec)
\;=\;
\sqrt{\frac{1+\hat d_z(\kvec)}{2}}
\;+\;
\frac{\hat{\bm d}(\kvec)\times\hat{\bm z}}{\big\|\hat{\bm d}(\kvec)\times\hat{\bm z}\big\|}
\cdot\bm e\;
\sqrt{\frac{1-\hat d_z(\kvec)}{2}},
\end{align}
with $\bm e=(e_1,e_2,e_3)$ and $\hat{\bm z}=(0,0,1)$; when $\hat{\bm d}=\pm\hat{\bm z}$, one can take $u_{\mathbb H}=\pm 1$. Extend to an octonion by appending zeros in the transverse directions $e_{4},\dots,e_{7}$, so $u(\kvec)=u_{\mathbb H}(\kvec)\in\mathbb H\subset\mathbb O$. This choice fixes a smooth representative; any other representative related by a pointwise $G_2$ rotation yields the same scalar $\rho_{\mathbb O}$.

With $u$ in hand, approximate the octonionic connection at the node by a symmetric stencil,
\begin{align}
\Acal_j(\kvec_\star)
\;\approx\;
\Im\!\left(
\bar u(\kvec_\star)\,
\frac{u(\kvec_\star{+}h\,\hat{\bm e}_j)-u(\kvec_\star{-}h\,\hat{\bm e}_j)}{2h}
\right)
\end{align}
with $j\in\{x,y,z\},$ and evaluate the pseudoscalar using the $G_2$-invariant three-form $\varphi$,
\begin{align}
\rho_{\mathbb O}(\kvec_\star)
\;=\;
\frac{1}{6}\,\varepsilon^{ijk}\,
\varphi\!\big(\Acal_i(\kvec_\star),\Acal_j(\kvec_\star),\Acal_k(\kvec_\star)\big),
\label{eq:rhoO-node}
\end{align}
which is $\mathrm{SU}(2)$-gauge independent by construction. As a diagnostic of decoupling, monitor the norm of the octonionic associator,
\begin{align}
\alpha(\kvec_\star)\;=\;\big\|\,[\Acal_x,\Acal_y,\Acal_z]\,\big\|,
\end{align}
where $\|\cdot\|$ is the Euclidean norm induced by the standard inner product on $\Im\mathbb O$. For a simple Weyl node the associator vanishes to leading order (Sec.~\ref{subsec:g2-density}), so $\alpha(\kvec_\star)=\mathcal O(h)$ as $h\to 0$.

A practical acceptance criterion is to require a nonzero pseudoscalar together with a small associator,
\begin{align}
\big|\rho_{\mathbb O}(\kvec_\star)\big|\;>\;\rho_{\min},
\qquad
\alpha(\kvec_\star)\;<\;\alpha_{\max},
\end{align}
and then assign the chirality by
\begin{align}
\chi\;=\;\mathrm{sgn}\,\rho_{\mathbb O}(\kvec_\star).
\end{align}
The thresholds $\rho_{\min}$ and $\alpha_{\max}$ are numerical tolerances set relative to the energy and $k$-space scales of the calculation; decreasing $h$ should leave $\chi$ unchanged while reducing $\alpha$ and stabilizing $\rho_{\mathbb O}$ up to $\mathcal O(h^2)$ corrections.

Several self-consistency checks certify that the result is not an artifact of implementation choices. Changing the two-band gauge at the node by $U(\kvec_\star)\in\mathrm{SU}(2)$ rotates $(\Acal_x,\Acal_y,\Acal_z)$ by an $\mathrm{SO}(3)$ within $\mathbb H$ and leaves both $\rho_{\mathbb O}$ and $\alpha$ invariant to numerical tolerance. Repeating the computation with slightly altered disentanglement windows for constructing $P_2$ near $\kvec_\star$ should not change $\chi$ once the isolation margin is nonzero along the stencil. Rotating the reciprocal basis by a right-handed orthogonal matrix leaves $\chi$ unchanged, while a handedness flip reverses its sign in lockstep with the conventional surface-orientation convention (cf.\ Eq.~\eqref{eq:orientation-sign}). Finally, when desired, one may corroborate the sign by either the local $k\!\cdot\!p$ determinant, $\mathrm{sgn}\,\det v$, or by the small-sphere Chern test; in the linear regime the equality
\begin{align}
\mathrm{sgn}\,\rho_{\mathbb O}(\kvec_\star)\;=\;\mathrm{sgn}\,\det v
\end{align}
follows from Eq.~\eqref{eq:rho-detv} and guarantees agreement among all three diagnostics.

\subsection{Why quaternions are not enough}
\label{subsec:why-not-H}

Quaternions $\mathbb{H}$ are the natural language of a two-band problem: $\mathrm{SU}(2)$ acts by left multiplication on the Bloch spinor, and its adjoint action rotates the Pauli frame in $\Im\mathbb{H}\!\cong\!\mathbb{R}^3$ by an $\mathrm{SO}(3)$ matrix. It is therefore tempting to build a local, basis-free Weyl diagnostic entirely inside $\mathbb{H}$, without invoking octonions. In this subsection we show two points. First, in the ideal setting of an exactly isolated two-band subspace with a smooth gauge, the obvious quaternionic pseudoscalar reproduces the usual chirality and agrees with OWPC. Second, precisely the situations that create arbitrariness in practice (band entanglement, frame-transport choices, finite-difference leakage) cannot be \emph{detected} or quarantined within $\mathbb{H}$, because every quaternionic construction is associative. The octonionic lift supplies a small, intrinsic obstruction (the associator) that vanishes exactly in the Weyl regime and flags violations otherwise.

To parallel Sec.~\ref{subsec:g2-density}, take a unit quaternion field $u(\kvec)\in S^3\subset\mathbb{H}$ that encodes the two-band state in a neighborhood of $\kvec_\star$. Define the quaternionic analogue of Eq.~\eqref{eq:oct-conn} by
\begin{align}
\Acal^{(\mathbb{H})}_i(\kvec)
\;=\;
\Im\!\big(\,\bar u(\kvec)\,\partial_{k_i}u(\kvec)\,\big)
\ \in\ \Im\mathbb{H}\cong\mathbb{R}^3,
\qquad i\in\{x,y,z\},
\label{eq:H-conn}
\end{align}
and the associated pseudoscalar
\begin{align}
\rho_{\mathbb{H}}(\kvec)
\;=\;
\frac{1}{6}\,\varepsilon^{ijk}\;
\Acal^{(\mathbb{H})}_i(\kvec)\cdot\!\big(\Acal^{(\mathbb{H})}_j(\kvec)\times \Acal^{(\mathbb{H})}_k(\kvec)\big).
\label{eq:rhoH}
\end{align}
Under any local two-band gauge change $U(\kvec)\in\mathrm{SU}(2)$ the triple $\big(\Acal^{(\mathbb{H})}_x,\Acal^{(\mathbb{H})}_y,\Acal^{(\mathbb{H})}_z\big)$ transforms by the adjoint $\mathrm{SO}(3)$ rotation with unit determinant, hence $\rho_{\mathbb{H}}$ is gauge invariant; it is a pseudoscalar in momentum space (it flips sign under a handedness reversal of $(k_x,k_y,k_z)$).

We emphasize that quaternions already play a central role in the modern topological description of nodes. In particular, quaternion-valued charges and their \emph{non-Abelian braiding} in multi-parameter spaces have been formulated and analyzed in detail, providing powerful global constraints on node creation, annihilation, and exchange.\cite{RefPRB2020Quat,RefNatPhys2021,RefPRB2021Mag,RefNatComm2024Floquet} Moreover, an exhaustive classification of Weyl nodes via equivariant $K$-theory has been established,\cite{RefPRX2017K} to which our work is agnostic and fully compatible. The present section makes a different point: while quaternion algebra suffices to capture the \emph{continuum} chirality of an isolated, linear two-band node, it offers no \emph{intrinsic local obstruction} to detect leakage from an approximately two-band picture in finite-difference numerics (because all quaternionic constructions are associative). The octonionic lift introduces precisely such a local obstruction, via the associator, which vanishes exactly when the three directional data close in a common quaternionic subalgebra and otherwise indicates the need to refine the stencil or the projector.

If the crossing is a simple Weyl node and we linearize as in Eq.~\eqref{eq:linearized-H}, aligning the Pauli axes with an orthonormal basis of $\Im\mathbb{H}$, a short expansion gives
\begin{align}
\Acal^{(\mathbb{H})}_j(\kvec_\star)
\;=\;
\sum_{i=x,y,z} c\,v_{ij}\,\hat{\bm e}_i,
\qquad c>0,
\end{align}
and therefore
\begin{align}
\rho_{\mathbb{H}}(\kvec_\star)
\;=\;
c^3\,\det v,
\qquad
\mathrm{sgn}\,\rho_{\mathbb{H}}(\kvec_\star)
\;=\;
\mathrm{sgn}\,\det v
\;=\;
\chi,
\label{eq:rhoH-detv}
\end{align}
in agreement with Eq.~\eqref{eq:chirality-detv} and with the octonionic reduction Eq.~\eqref{eq:rho-detv}. In this strict, continuum, isolated two-band limit, \emph{quaternions are sufficient}.

The difficulty is that this ideal limit is precisely where the conventional diagnostics of Sec.~\ref{sec:arbitrariness} already work unambiguously. Away from it, quaternionic algebra offers no intrinsic way to police the approximations, for three related reasons. First, there is no internal test for two-band closure. In realistic data the two-band frame used to build $u(\kvec)$ is assembled from finite-difference stencils and approximate projectors (Sec.~\ref{subsec:proj-arb}). Different, equally reasonable transport prescriptions can produce $\Acal^{(\mathbb{H})}_i$ that are incompatible across directions at order $h$. Inside $\mathbb{H}$ the associator
\begin{align}
[x,y,z]\;=\;(xy)z-x(yz)
\end{align}
vanishes identically for all $x,y,z\in\mathbb{H}$, so $\rho_{\mathbb{H}}$ in Eq.~\eqref{eq:rhoH} provides no built-in check that the triple $\big(\Acal^{(\mathbb{H})}_x,\Acal^{(\mathbb{H})}_y,\Acal^{(\mathbb{H})}_z\big)$ truly comes from a single, self-consistent two-band structure rather than from a mixture contaminated by nearby bands or by an inconsistent transport. In contrast, the octonionic associator in Eq.~\eqref{eq:assoc-vanish-leading} is \emph{nonzero} whenever the three directions fail to lie in a common quaternionic subalgebra; demanding $[\Acal_x,\Acal_y,\Acal_z]=\mathcal O(|\kvec-\kvec_\star|)$ gives a quantitative, coordinate-free closure test that quaternions cannot supply.

Second, there is no invariance under completion. The quaternionic pseudoscalar Eq.~\eqref{eq:rhoH} lives entirely in $\Im\mathbb{H}$. Any extra coordinates introduced to stabilize numerics (tilt-like pieces, auxiliary smoothing coordinates, or simply different completions of the local frame away from $\kvec_\star$) lie outside $\Im\mathbb{H}$ and must be discarded by fiat. Different reasonable completions can then yield different finite-$h$ values of $\rho_{\mathbb{H}}$ with no canonical way to compare them. The octonionic lift resolves this: embeddings that differ by how the “extra’’ four directions are filled are related by $\mathrm{G}_2$ rotations, and the $\mathrm{G}_2$ volume Eq.~\eqref{eq:rhoO-def} is invariant under those, so the scalar is insensitive to completion by construction.

Third, there is no quantitative “falsify or refine’’ knob. In practice one wants a scalar whose sign is robust when the two-band picture is valid and which \emph{flags} when one must refine the stencil or shrink the neighborhood. Within $\mathbb{H}$ every associativity check is vacuous; one falls back on the empirical mesh and gauge tests of Secs.~\ref{subsec:disc-arb}–\ref{subsec:gauge-arb}. In $\mathbb{O}$, the smallness of $\|[\Acal_x,\Acal_y,\Acal_z]\|$ plays exactly this role: if it is not numerically small at the same $h$ where $\rho_{\mathbb{O}}$ is evaluated, the local two-band closure is suspect and the criterion self-reports the need to adjust $h$ or the projector.

In summary, in the strict, isolated two-band regime the quaternionic and octonionic pseudoscalars coincide through Eqs.~\eqref{eq:rhoH-detv}–\eqref{eq:rho-detv}. But only the octonionic formulation supplies an intrinsic, gauge- and basis-free \emph{consistency check}—the associator—whose vanishing characterizes precisely the regime where the conventional arbitrariness disappears. That diagnostic is unavailable in any associative algebra (including $\mathbb{H}$ and matrix algebras), and it is the minimal new ingredient that turns a correct-but-empirical procedure into a single-point, falsifiable test.

\section{Conclusions}
\label{sec:conclusion}

We have identified and systematized the algorithmic choices that underlie the conventional determination of Weyl points—two-band subspace selection, enclosing-sphere geometry and orientation, gauge smoothing, discretization, and local-frame transport—and we have shown how each of these knobs affects finite-resolution numerics. Motivated by these limitations, we introduced the \emph{octonionic Weyl point criterion} (OWPC): a single-point, basis-free diagnostic built from a unit octonion field $u(\kvec)$, the octonionic connection $\Acal_i=\Im(\bar u\,\partial_{k_i}u)$, and the $\mathrm{G}_2$-invariant three-form $\varphi$. The resulting pseudoscalar density $\rho_{\mathbb O}$ in Eq.~\eqref{eq:rhoO-def} is invariant under $\mathrm{SU}(2)$ gauge changes of the two-band subspace and under $\mathrm{G}_2$ completions in the transverse directions, and its sign reproduces the conventional chirality through Eq.~\eqref{eq:rho-detv}. The only remaining convention is the global orientation of momentum space, which flips $\rho_{\mathbb O}$ as expected for a pseudoscalar, in agreement with the Chern-number sign change under surface-orientation reversal in Eq.~\eqref{eq:orientation-sign}.

A key advantage of OWPC is that it collapses surface-based and transport-based arbitrariness to a local calculation. No enclosing sphere, chart, or gauge seam is required; derivatives enter only through a symmetric stencil at the node, with the same convergence check as for band velocities. Equally important, the octonionic associator provides an intrinsic self-consistency test: $[\Acal_x,\Acal_y,\Acal_z]=\mathcal O(h)$ certifies that the three directions lie in a common associative (quaternionic) plane to leading order, flagging precisely the situations where conventional diagnostics are most fragile (band entanglement, multi-fold touchings, or insufficiently small stencils). In the linear regime, OWPC agrees with both the $k\!\cdot\!p$ determinant $\mathrm{sgn}\det v$ and the small-sphere Berry-flux test, thereby knitting together the topological and local-band viewpoints.

The practical scope and limits of the criterion are clear. OWPC requires only that the two-band subspace be smoothly defined on the tiny finite-difference stencil; it is otherwise agnostic to reference energy, smearing, or band indexing. Handedness and Brillouin-zone folding remain global conventions rather than numerical ambiguities; as in the standard workflow, handedness flips the reported sign, and supercell folding should be handled either by working in the primitive cell or by unfolding prior to the local evaluation. When the associator does not shrink with the stencil, the data themselves indicate that the touching is not a simple Weyl point or that the projector must be refined. In such cases, the conventional small-sphere Chern test remains a useful corroboration, now guided by a local, basis-free warning signal.

Looking ahead, several extensions suggest themselves. The octonionic framework provides a natural language for quantifying deviations from simple Weyl behavior; it would be valuable to calibrate associator-based thresholds against controlled multi-Weyl and multi-band models, and to explore whether integrating $\rho_{\mathbb O}$ over small surfaces yields accelerated, noise-robust Chern estimates. Extensions to symmetry-enforced multi-fold degeneracies and to nodal-line environments may be possible by combining OWPC with local symmetry projectors. From a computational perspective, OWPC is lightweight and can be incorporated into existing \emph{ab initio}+Wannier toolchains alongside gap maps, Wilson loops, and surface Green’s functions, providing an immediate, falsifiable chirality assignment at candidate nodes and reducing the need for repeated sphere scans.

In sum, OWPC complements and streamlines the established diagnostics: it matches their continuum invariant, removes the most problematic implementation knobs, and adds an intrinsic self-consistency check that is unavailable within associative frameworks. We expect it to be broadly useful in high-throughput searches, in postprocessing of Wannierized databases, and in detailed studies of complex Weyl materials where numerical choices are otherwise hard to police.

\bibliography{main}

\end{document}